\documentclass[a4paper,11pt]{article}

\usepackage{pos}
\usepackage{subcaption}
\usepackage{lipsum} 

\title{Mechanical design of the optical modules intended for IceCube-Gen2}

\ShortTitle{Mechanical design of the optical modules intended for IceCube-Gen2}

\author{The IceCube-Gen2 Collaboration \\{\normalsize \normalfont(a complete list of authors can be found at the end of the proceedings)}\\}

\emailAdd{yuya.makino@icecube.wisc.edu}

\abstract{

IceCube-Gen2 is an expansion of the IceCube neutrino observatory at the South Pole that aims to increase the sensitivity to high-energy neutrinos by an order of magnitude. To this end, about 10,000 new optical modules will be installed, instrumenting a fiducial volume of about 8~km$^3$. Two newly developed optical module types increase IceCube’s current sensitivity per module by a factor of three by integrating 16 and 18 newly developed four-inch PMTs in specially designed 12.5-inch diameter pressure vessels. Both designs use conical silicone gel pads to optically couple the PMTs to the pressure vessel to increase photon collection efficiency. The outside portion of gel pads are pre-cast onto each PMT prior to integration, while the interiors are filled and cast after the PMT assemblies are installed in the pressure vessel via a pushing mechanism. This paper presents both the mechanical design, as well as the performance of prototype modules at high pressure (70~MPa) and low temperature (-40~${}^\circ$C), characteristic of the environment inside the South Pole ice.

\vspace{4mm}
{\bfseries Corresponding authors:}
Yuya Makino$^{1*}$ \\ 
{$^{1}$ \itshape  Dept. of Physics and Wisconsin IceCube Particle Astrophysics Center, University of Wisconsin–Madison, Madison, WI 53706, USA}\\ [4mm]
$^*$ Presenter

\ConferenceLogo{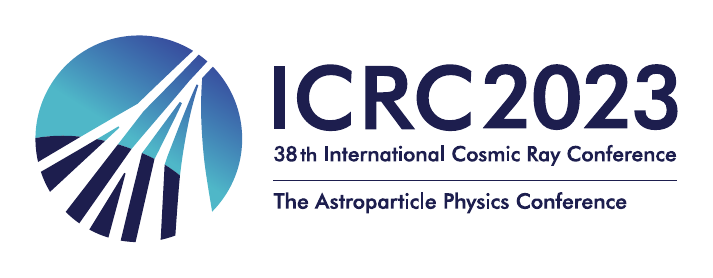}

\FullConference{The 38th International Cosmic Ray Conference (ICRC2023)\\ 26 July -- 3 August, 2023\\ Nagoya, Japan}
}

\begin{document}

\maketitle

\section{Introduction}\label{sec:intro}
 
IceCube-Gen2 \cite{ishihara:2023icrc} is a high energy extension of IceCube Neutrino Observatory \cite{Aartsen:2016nxy} at the South Pole.
The planned optical array of the IceCube-Gen2 consists of 120 new strings, instrumented between depths 1.35 and 2.7~km, with 240~m inter-string spacing. 
Each string has 80 Optical Modules with 17~m separation.
About 8~km$^3$ of the detector volume will be covered by 9,600 new Optical Modules. 
Due to the extended inter-string spacing compared to IceCube, 125~m, we aim to develop an Optical Module that has four-times better sensitivity compared to IceCube Digital Optical Module (DOM) \cite{gen2tdr}.

Achieving this high sensitivity is possible by installing a larger number of small diameter Photomultiplier Tubes (PMTs) in a larger pressure vessel, KM3NeT DOM \cite{Aiello_2022}, for example.  
However, unlike other open water neutrino telescopes, the enormous cost of ice drilling at the South Pole limits the size of the pressure vessel. The requirement on the diameter of the module is 12.5 inch, even smaller than that of IceCube (13 inch).  

The biggest challenge of the development of the Optical Module for IceCube-Gen2 is to achieve extremely high per-module sensitivity with a very compact Optical Module design. 
That unavoidably leads the mechanical design of the module to be complex and tightly packed.
Yet the module candidate need to satisfy severe mechanical requirements before and after the deployment; transportation to the Geographic South Pole, low temperature ($-$40~${}^\circ$C), and high pressure (70~MPa) during the refreezing phase after the deployment.
Moreover, the modules have to be extremely robust as they will be physically inaccessible after the deployment. 
IceCube DOMs have shown quite high reliability after over a decade of operation; 98\% of the deployed modules are still taking data.
The new Optical Module needs to be carefully discussed about not only the optical performances, but also its production feasibility and mechanical strength. 

%

\section{IceCube-Gen2 DOM design candidates}\label{sec:gen2dom}

Initial conceptual studies have concluded that a four-inch PMT is the best choice to achieve the highest sensitivity for the Optical Module with a 12.5 inch pressure vessel. 
The reason is that larger PMTs make a back-to-back layout practically impossible while covering all directions with PMTs, and smaller PMTs lead to inefficient packing in the pressure vessel. 
Two PMT vendors, Hamamatsu Photonics K.K. and North Night Vision Technology developed 4 inch PMTs with very short total length (106~mm) without compromising the optical performance \cite{dittmer:2023icrc}.

Naulitus GmbH and Okamoto Glass made custom-design pressure vessels and they are confirmed to meet the requirement, 70~MPa, at pressure testing facilities. 
The diameter and height of each vessel are 318~mm $\times$ 540~mm and 312~mm $\times$ 444~mm, respectively. 
Both vessels have elongated shapes to maximize the capacity inside the vessels to store as many four-inch PMTs as possible. 
Properties of both glass, such as UV transmittance and radioactivity, have been studied through the Optical Module developments for IceCube-Upgrade; D-Egg \cite{Abbasi_2023} and mDOM \cite{anderson2021design}.

\begin{figure}
     \centering
     \begin{subfigure}[b]{0.49\textwidth}
         \centering
         \includegraphics[width=0.95\textwidth]{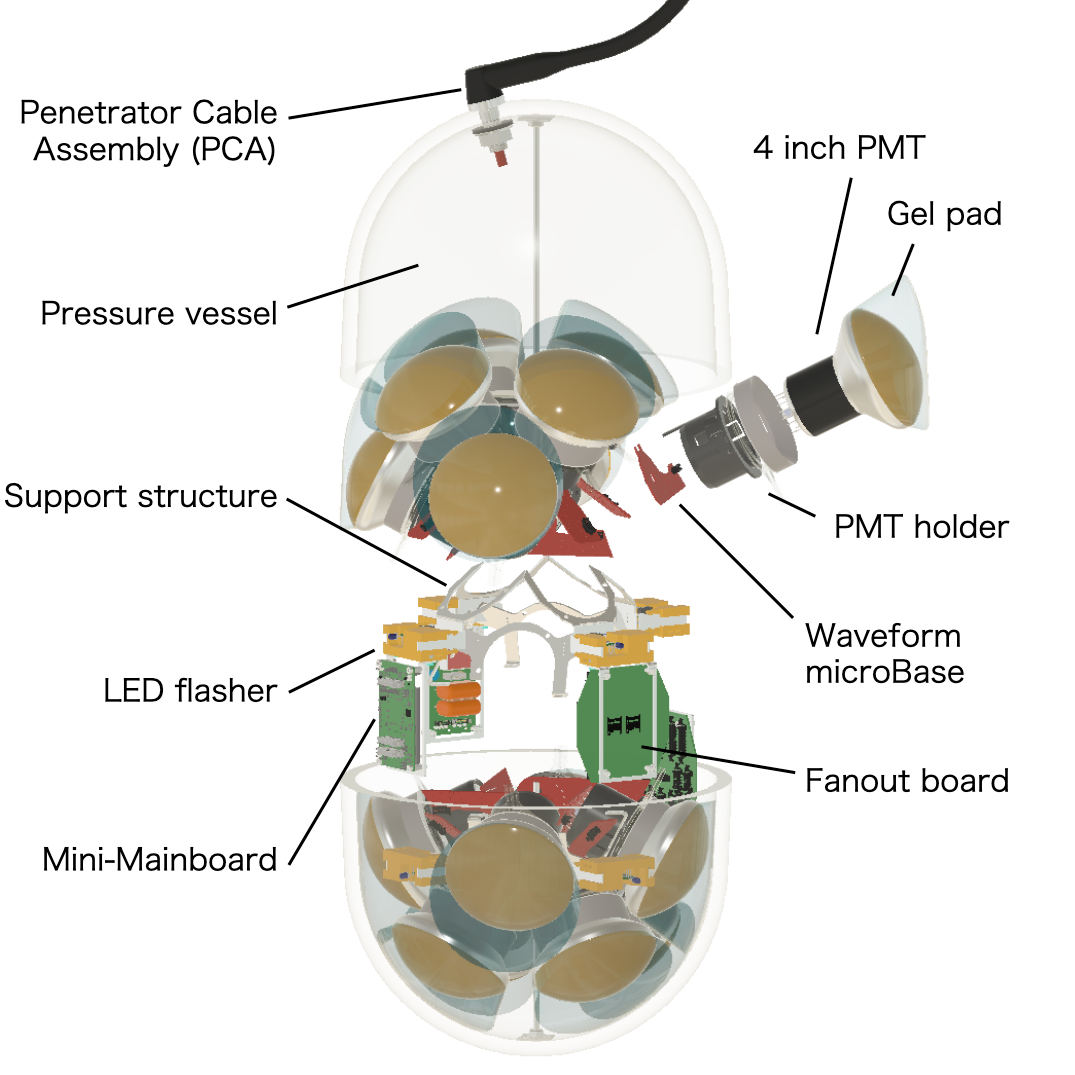}         
         \caption{16 PMT model (exploded view)}
         \label{fig:model16}
     \end{subfigure}
     \hfill
     \begin{subfigure}[b]{0.49\textwidth}
         \centering
         \includegraphics[width=0.95\textwidth]{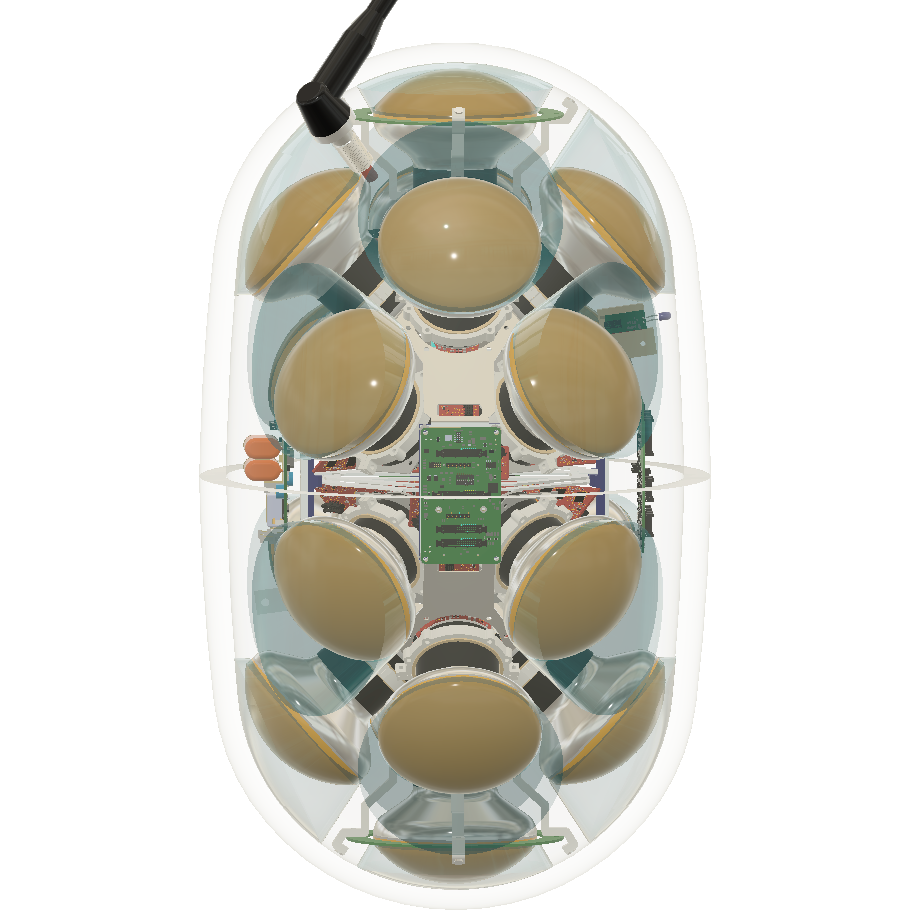}         
         \caption{18 PMT model}
         \label{fig:model18}
     \end{subfigure}
     \hfill
    \caption{Schematic designs of Gen2 DOM 16 (left) and 18 PMT (right) models. The images are not to scale. Sealing tapes and harnesses are not shown in these images.}
    \label{fig:gen2doms}
\end{figure}

Figure~\ref{fig:gen2doms} shows the schematic design views of the design candidates of the IceCube-Gen2 DOM; 16 and 18 PMT models. 
Each PMT assembly consists of a four-inch PMT, ``gel pad", PMT holder, and Waveform microBase (wuBase) \cite{Basu_2021}.
Gel pads, conical-shaped optical gel cones, couples the PMT to the pressure vessel, which is explained in details in Sec.\ref{sec:gelpad}.
The wuBase is a Cockcroft–Walton HV base with data taking features, reading the anode and 8th dynode signals with 2-channel 60 mega samples per second (MSPS) ADC \cite{griffin:2023icrc}. 
PMT assemblies are mounted on a support structure that is made of sheet metal.
The support structure also holds central electronics boards; Fanout boards and Mini-Mainboards that are responsible for sending commands and collecting data to/from each PMT and communication with the surface through Penetrator Cable Assembly (PCA).
For more details on the Gen2 DOM electronics, see \cite{griffin:2023icrc}.
The top and bottom PMT assemblies of the 18 PMT model have central shafts that hold the support structures and components attached to them in each hemisphere. 
Although the 16 PMT model has a similar central shaft, directly attached to the pressure vessel, it is for the alignment during assembly. The gel pads hold all components in the vessel after the integration. 

The upper and lower hemispheres are independently assembled and put together with a mechanical spring. Only two ribbon cables crossing the hemispheres electrically connect them.  
The hemispheres are hold together by vacuuming the vessel to a half atmospheric pressure and sealed with butyl rubber and a layer of protection tape at the equator.
The vacuuming and sealing procedure follow the one established for IceCube DOM \cite{Aartsen:2016nxy}. 
Overall, both models share the same concept and components, but use slightly different approaches on the PMT holders.

Simulation demonstrates that the 16 and 18 PMT models achieve more than four times better sensitivities compared to IceCube DOM \cite{gen2tdr}. The improvement is rather significant at the horizontal directions which is critical for the horizontally-sparse array of IceCube-Gen2.

\section{Gel pad}\label{sec:gelpad}

mDOM \cite{anderson2021design} and KM3NeT \cite{Aiello_2022} have successfully demonstrated the use of multiple relatively-small PMTs in Optical Module designs for neutrino telescopes. 
In case of mDOM, two dome-shaped support structures hold 24 three-inch PMTs in a 356~mm diameter vessel.
The gap between the pressure vessel hemisphere and all internal structure held by the support structure is thoroughly filled with optical gel. 
The gel holds all internal components of each hemisphere after it cures. 
The support structure is manufactured by 3D printing due to its highly-complex shape. 
Similar technology was used in KM3Net DOM design, although their recent production uses injection molding for the support structure production \cite{Aiello_2022}.  

Considering the production size of IceCube-Gen2 Optical Modules, 9,6000 in total, components made of 3D printing is not preferable as its unit price basically does not lower for a larger scale production. Additionally, filling the gap between a vessel and the support structure requires significant amount of optical gel. This will lead higher per-module price and weight.  

An idea so-called ''gel pad" have been studied for the IceCube-Gen2 DOM design \cite{Basu_2021}.
Gel pads are conical-shaped optical gel that connect the photocathode area of PMTs and the inner surface of the vessel. Total reflection on the conical surface enhances the photodetection efficiency of the PMTs.
Similar ideas have been previously studied in other experiments (\cite{Hyper-Kamiokande:2016dsw}, for example).



For production, we cast gel onto a PMT prior to the integration.
A two-pieces vacuum-formed thin mold made of polypropylene or high-impact polystyrene (HIPS) covers the photocathode area of a PMT. 

\begin{figure}
     \centering
     \begin{subfigure}[b]{0.49\textwidth}
         \centering
         \includegraphics[width=1.0\textwidth]{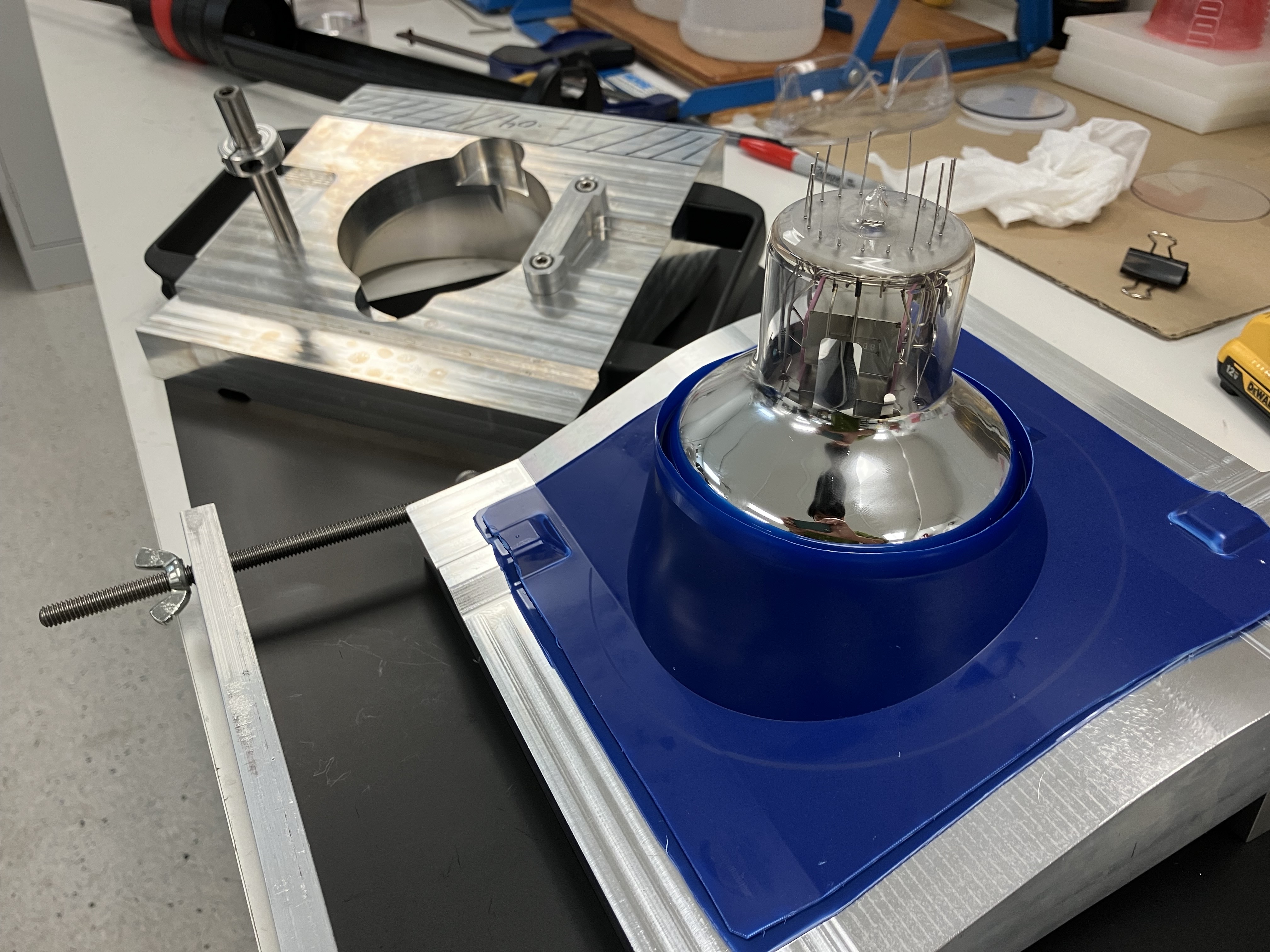}         
         \caption{HIPS molds and a four-inch PMT mounted on jigs.}
         \label{fig:gelpadmold}
     \end{subfigure}
     \hfill
     \begin{subfigure}[b]{0.49\textwidth}
         \centering
         \includegraphics[width=0.75\textwidth]{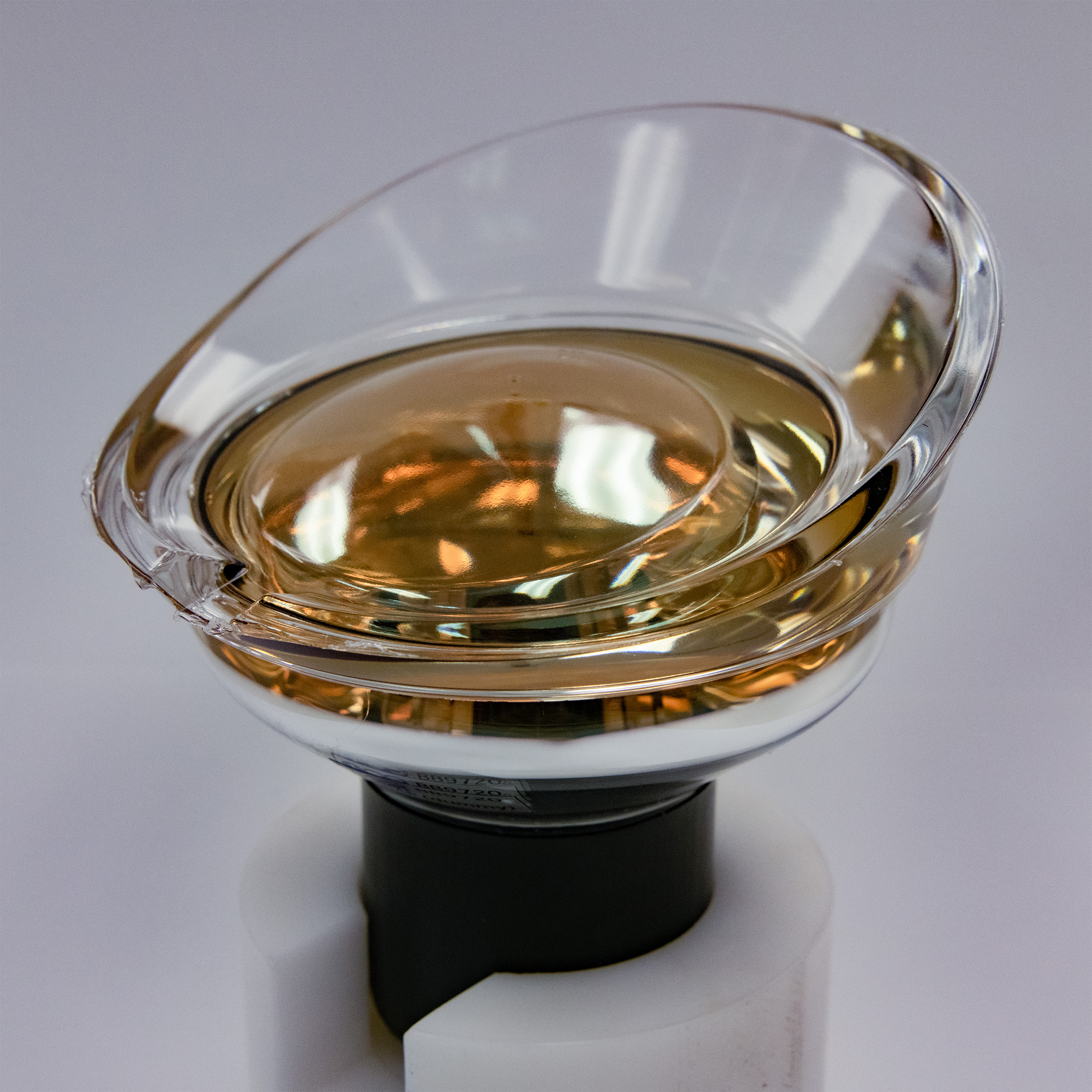}         
         \caption{Casted gel-pad on a 4 inch PMT.}
         \label{fig:gelpadgelpad}
     \end{subfigure}
     \hfill
    \caption{(Left) A four-inch PMT covered by two thin HIPS molds are fixed in a metal jig during the casting. (Right) Improved molds give high surface quality and allows to de-mold easily without damaging the pads.}
    \label{fig:gelpad}
\end{figure}

The PMT and molds are tightly fixed in a jig during the casting process until the gel cures after a day. 
The molds holding PMTs after casting is shown in Fig~\ref{fig:gelpadmold}. 
The materials of the molds are carefully chosen from studies on the vacuum forming quality and ease of de-molding after casting to improve the efficiency of total reflection on the conical surface areas. 
The gel is ShinEsu X3547-HE that has been developed for the D-Egg and is known for high UV transmittance and large enough elasticity to accomodate the thermal compression down to $-$40~${}^\circ$C.

Figure~\ref{fig:gelpadgelpad} shows a four-inch PMT after the casting. 
The rim shape, 5~mm width, is designed to match the inner surface of the pressure vessel. 
An opening seen on the left is intentionally left open in order to back-fill additional gel into a cavity at the center.
In-place casting of the gel pads, for example, is practically not feasible because of de-molding in the limited space of the highly packed layouts inside the module.  
The two-step process, pre-casting and back-filling, was developed through in-lab iterations, and was found to give the best results.

\section{Integration and mechanical testings}\label{sec:integration}

Each PMT holder has a pushing mechanism to move itself towards the pressure vessel surface. 
After placing the pressure vessel hemisphere on the assembly jig, we activate the pushing mechanism and the rims of each PMT assembly touch the internal surface of the vessel. We apply a thin layer of uncured gel on the rim before pressing it to the vessel to ensure the sealing between the vessel and wait for a day before moving to the next step.
    
\begin{figure}
     \centering
     \begin{subfigure}[b]{0.49\textwidth}
         \centering
         \includegraphics[width=0.6\textwidth]{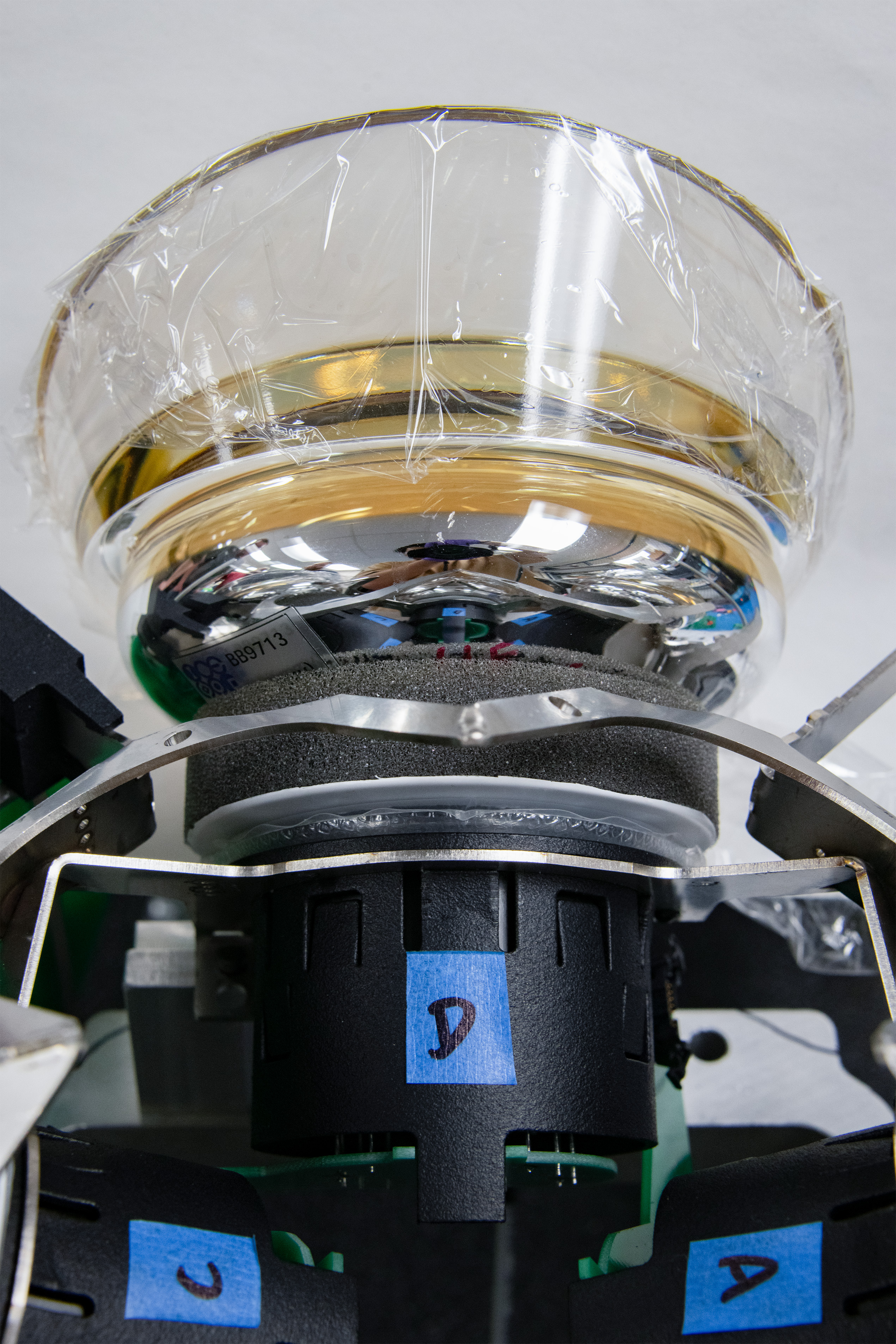}         
         \caption{16 PMT model}
         \label{fig:push16}
     \end{subfigure}
     \hfill
     \begin{subfigure}[b]{0.49\textwidth}
         \centering
         \includegraphics[width=0.9\textwidth]{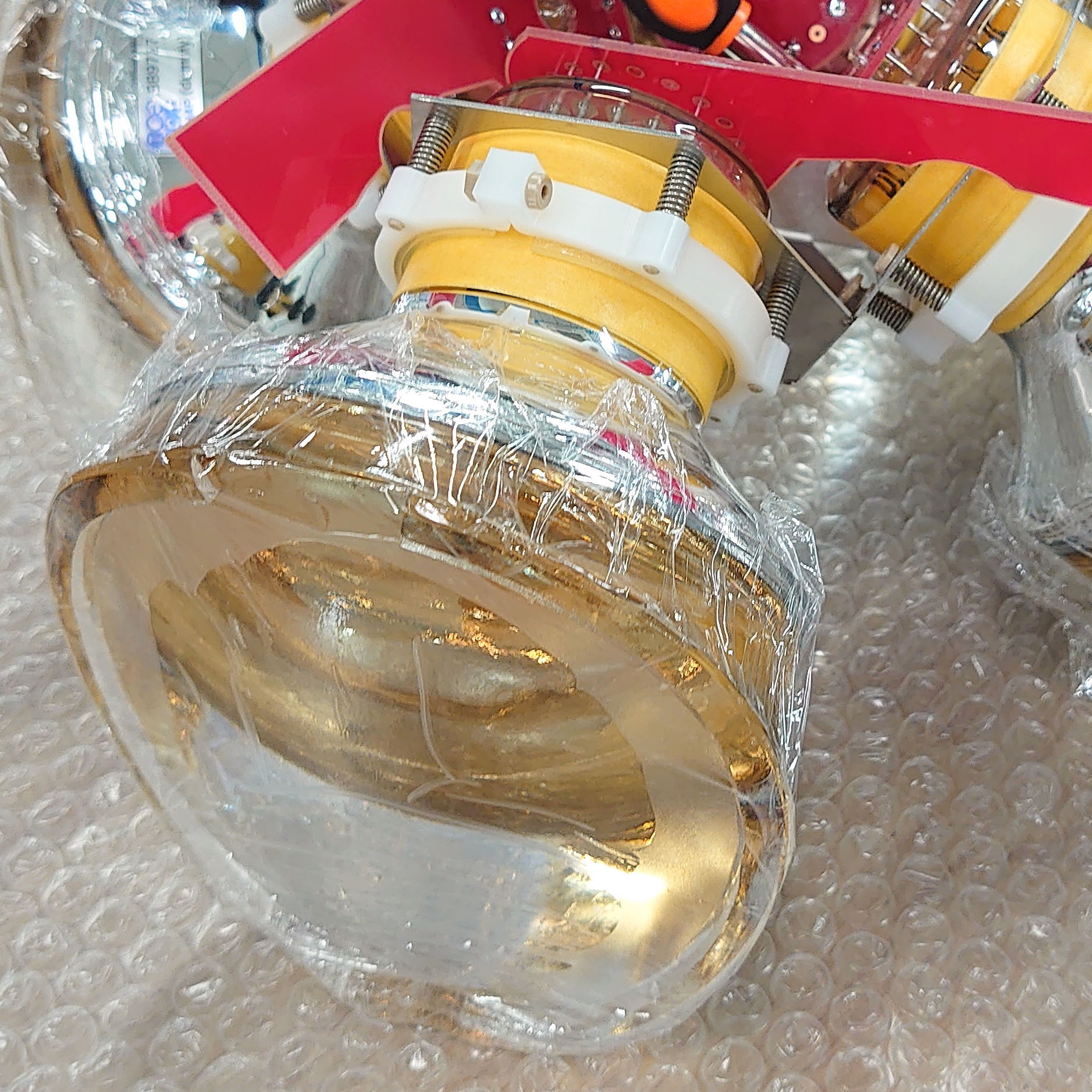}         
         \caption{18 PMT model}
         \label{fig:push18}
     \end{subfigure}
     \hfill
    \caption{Pushing mechanism of the 16 (left) and 18 PMT (right) models. The gel pads are covered by thin plastic films for protection before the integration.}
    \label{fig:pusing-mechanisms}
\end{figure}

Figure~\ref{fig:pusing-mechanisms} shows the pushing mechanism of both the 16 and 18 PMT models. Although the two models share the same sub-components, different approaches are still being explored to maximize effectiveness.  
For the 16 PMT model, a custom-made inflating tube made of thin plastic sheets and open-cell form surround the PMT neck (Fig.~\ref{fig:push16}). By injecting silicone rubber into the inflating tube, it expands pushing the PMT about 5~mm towards the inner wall of the pressure vessel. 
The 18 PMT model adopts mechanical springs installed between the support structure and a plastic ring attached to each PMT neck (Fig.~\ref{fig:model18}). The springs are tightened before integration and gradually released until the gel pad attaches to the inner wall of the vessel during the integration. 
Both mechanisms additionally act as absorbers, as a 3-4~mm horizontal compression of the pressure vessel from surrounding ice at a depth of 1-3~km is expected during deployment.

\begin{figure}
    \centering
    \includegraphics[width=0.8\textwidth]{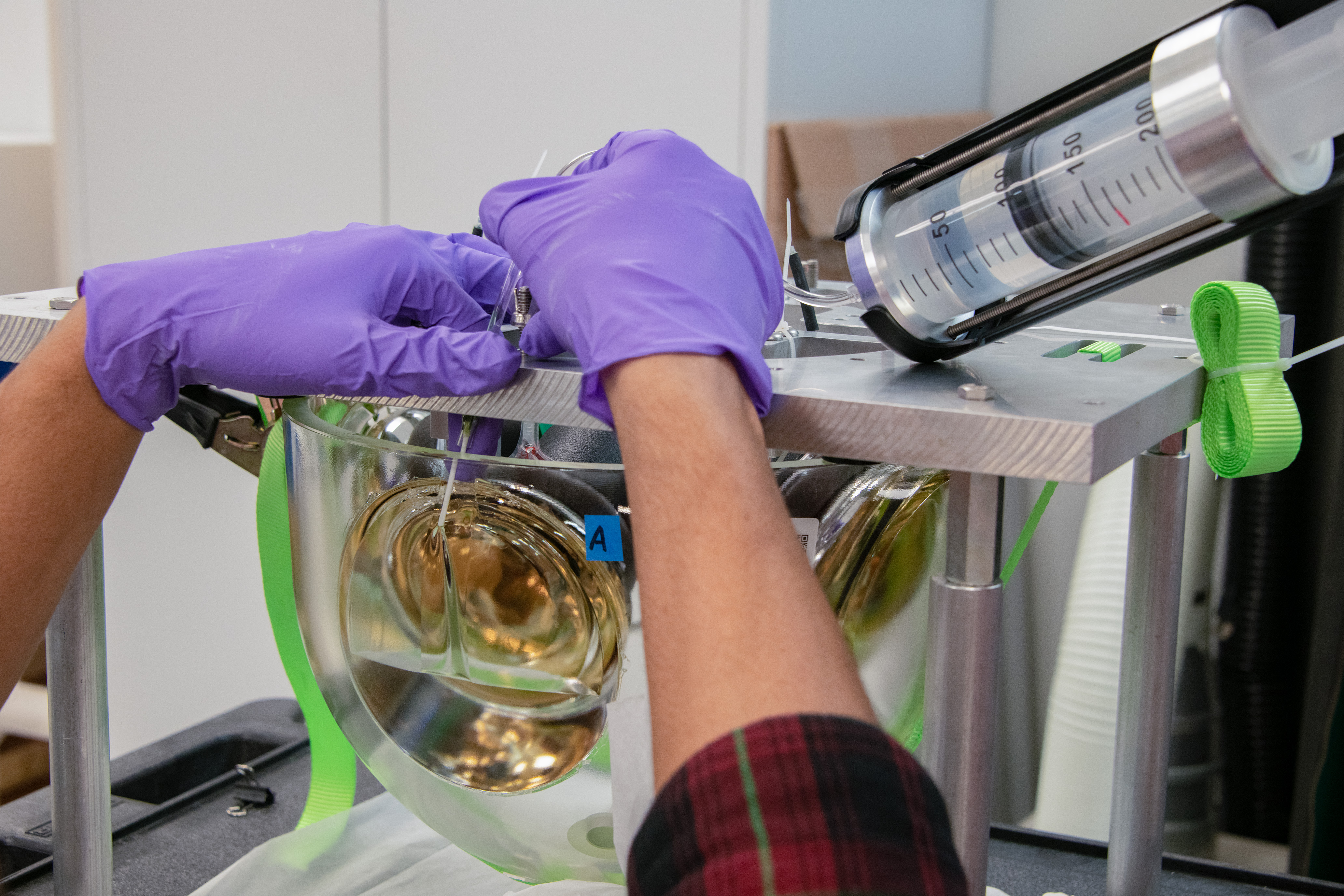}     
    \caption{Back-filling the gel into a cavity of one of the gel pads. A simple setup, a thin tube and syringe, is used for the prototyping, which will be replaced with a dispensing machine for the module mass production phase. The metal base holds the hemisphere during the assembly.}
    \label{fig:backfilling}
\end{figure}

\begin{figure}
     \centering
     \begin{subfigure}[b]{0.49\textwidth}
         \centering
         \includegraphics[width=\textwidth]{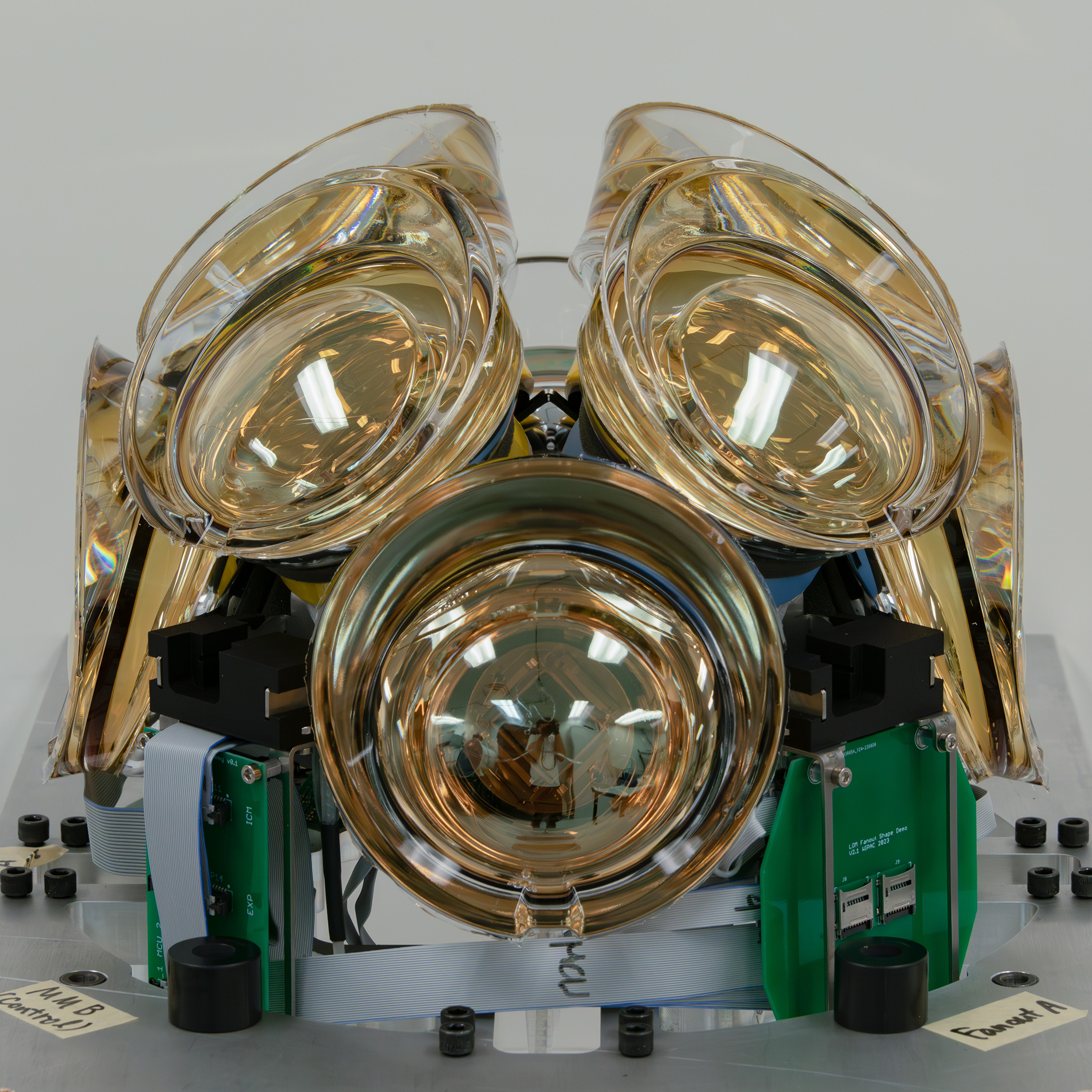}         
         \caption{16 PMT model before back filling}
         \label{fig:proto16pmt}
     \end{subfigure}
     \hfill
     \begin{subfigure}[b]{0.49\textwidth}
         \centering
         \includegraphics[width=\textwidth]{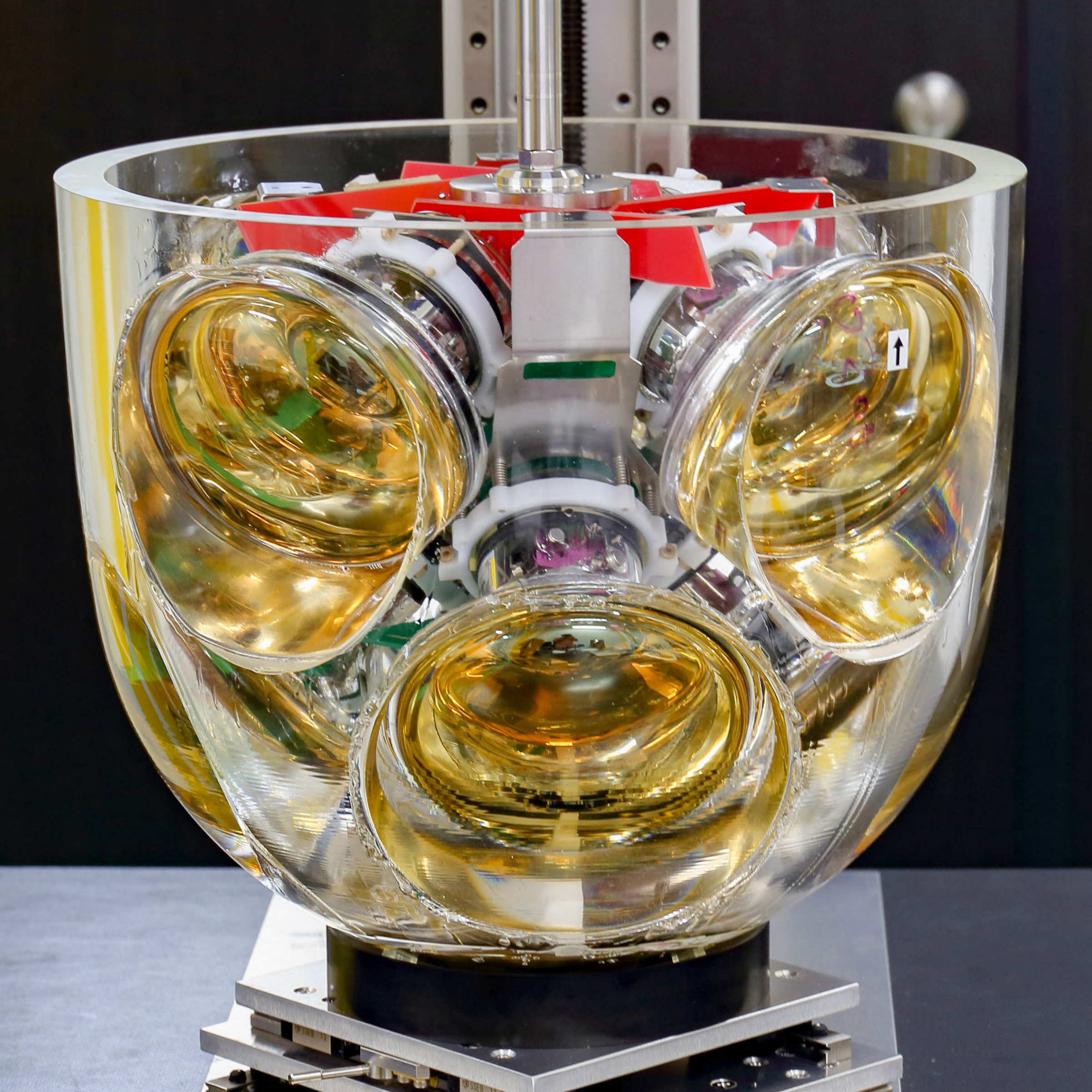}         
         \caption{18 PMT model after back filling}
         \label{fig:proto18pmt}
     \end{subfigure}
     \hfill
    \caption{Integration prototype models}
    \label{fig:tada}
\end{figure}

Once the rims of the gel pads are securely sealed, an additional is poured into the cavities from the openings of each gel pad (Fig.~\ref{fig:backfilling}).  
After completing the back-filling process, the entire hemisphere is placed in a vacuum chamber to remove the bubbles remained in the gel pads. These bubbles can escape from the opening that locates at the top during this procedure as seen in Fig.~\ref{fig:backfilling}. 
Figure~\ref{fig:tada} shows integration prototypes of 16 and 18 PMT models.  

Assembled prototypes are tested at low temperature.
We visually inspected all gel pads under $-$40~${}^\circ$C environment for more than a day. We observed no issues with the gel pad, such as cracks in the pads or delamination between the gel and PMTs or the gel and pressure vessel. 
We also tested the effect of high pressure on the internal structure with the 18 PMT model prototype at Japan Agency for Marine-Earth Science and Technology (JAMSTEC). 
No damages in the components due to the compression of the pressure vessel at 70~MPa were found after multiple testings.  
We plan to perform the same pressure test for the 16 PMT model in Summer 2023. 

\section{Summary and outlook}\label{sec:summary}
The next-generation optical module is under development for eventual deployment in IceCube-Gen2.
The current designs adopt 16 or 18 four-inch PMTs in a 12.5 inch vessel and shows 4 times better sensitivity compared to the IceCube DOM. 
The gel pads offer a unique approach to optically coupling the PMTs to the pressure vessel, reducing both the weight and cost without losing photon detection performance.
The casting technique of the gel pad is well established and the integration prototypes showed no issues at low temperature nor at high pressure. 
We aim to build a completely functional Optical Module with the new design in 2023 and plan to deploy 12 modules in an engineering and development section of the IceCube-Upgrade array in the 2025/26 South Pole summer season. 
This will serve as final verification of the optical module design, looking towards construction of the IceCube-Gen2 array in early 2030.


\bibliographystyle{ICRC}
\bibliography{references}

%

\clearpage

\section*{Full Author List: IceCube-Gen2 Collaboration}

\scriptsize
\noindent
R. Abbasi$^{17}$,
M. Ackermann$^{76}$,
J. Adams$^{22}$,
S. K. Agarwalla$^{47,\: 77}$,
J. A. Aguilar$^{12}$,
M. Ahlers$^{26}$,
J.M. Alameddine$^{27}$,
N. M. Amin$^{53}$,
K. Andeen$^{50}$,
G. Anton$^{30}$
% ...
\\
\\
% ...
$^{12}$ Universit{\'e} Libre de Bruxelles, Science Faculty CP230, B-1050 Brussels, Belgium \\
% ...
$^{17}$ Department of Physics, Loyola University Chicago, Chicago, IL 60660, USA \\
% ...
$^{22}$ Dept. of Physics and Astronomy, University of Canterbury, Private Bag 4800, Christchurch, New Zealand \\
% ...
$^{26}$ Niels Bohr Institute, University of Copenhagen, DK-2100 Copenhagen, Denmark \\
$^{27}$ Dept. of Physics, TU Dortmund University, D-44221 Dortmund, Germany \\
% ...
$^{30}$ Erlangen Centre for Astroparticle Physics, Friedrich-Alexander-Universit{\"a}t Erlangen-N{\"u}rnberg, D-91058 Erlangen, Germany \\
% ...
$^{47}$ Dept. of Physics and Wisconsin IceCube Particle Astrophysics Center, University of Wisconsin{\textendash}Madison, Madison, WI 53706, USA \\
% ...
$^{50}$ Department of Physics, Marquette University, Milwaukee, WI, 53201, USA \\
% ...
$^{53}$ Bartol Research Institute and Dept. of Physics and Astronomy, University of Delaware, Newark, DE 19716, USA \\
% ...
$^{76}$ Deutsches Elektronen-Synchrotron DESY, Platanenallee 6, 15738 Zeuthen, Germany  \\
$^{77}$ Institute of Physics, Sachivalaya Marg, Sainik School Post, Bhubaneswar 751005, India \\
% ...

\subsection*{Acknowledgements}

\noindent
The authors gratefully acknowledge the support from the following agencies and institutions:
USA {\textendash} U.S. National Science Foundation-Office of Polar Programs,
U.S. National Science Foundation-Physics Division,
U.S. National Science Foundation-EPSCoR,
Wisconsin Alumni Research Foundation,
Center for High Throughput Computing (CHTC) at the University of Wisconsin{\textendash}Madison,
Open Science Grid (OSG),
Advanced Cyberinfrastructure Coordination Ecosystem: Services {\&} Support (ACCESS),
Frontera computing project at the Texas Advanced Computing Center,
U.S. Department of Energy-National Energy Research Scientific Computing Center,
Particle astrophysics research computing center at the University of Maryland,
Institute for Cyber-Enabled Research at Michigan State University,
and Astroparticle physics computational facility at Marquette University;
Belgium {\textendash} Funds for Scientific Research (FRS-FNRS and FWO),
FWO Odysseus and Big Science programmes,
and Belgian Federal Science Policy Office (Belspo);
Germany {\textendash} Bundesministerium f{\"u}r Bildung und Forschung (BMBF),
Deutsche Forschungsgemeinschaft (DFG),
Helmholtz Alliance for Astroparticle Physics (HAP),
Initiative and Networking Fund of the Helmholtz Association,
Deutsches Elektronen Synchrotron (DESY),
and High Performance Computing cluster of the RWTH Aachen;
Sweden {\textendash} Swedish Research Council,
Swedish Polar Research Secretariat,
Swedish National Infrastructure for Computing (SNIC),
and Knut and Alice Wallenberg Foundation;
European Union {\textendash} EGI Advanced Computing for research;
Australia {\textendash} Australian Research Council;
Canada {\textendash} Natural Sciences and Engineering Research Council of Canada,
Calcul Qu{\'e}bec, Compute Ontario, Canada Foundation for Innovation, WestGrid, and Compute Canada;
Denmark {\textendash} Villum Fonden, Carlsberg Foundation, and European Commission;
New Zealand {\textendash} Marsden Fund;
Japan {\textendash} Japan Society for Promotion of Science (JSPS)
and Institute for Global Prominent Research (IGPR) of Chiba University;
Korea {\textendash} National Research Foundation of Korea (NRF);
Switzerland {\textendash} Swiss National Science Foundation (SNSF);
United Kingdom {\textendash} Department of Physics, University of Oxford.

\end{document}